\newcommand{\version}{April 2, 2011}
\documentclass[a4paper,twoside,11pt,pdftex]{article}
\usepackage[bindingoffset=0.4cm,textheight=22.5cm,hdivide={2.7cm,*,2.75cm}, vdivide={*,22cm,*}]{geometry}
\usepackage{amsmath,amsfonts,amssymb}
\IfFileExists{dsfont.sty}
	{\usepackage{dsfont}
         \let\mathbb=\mathds
         \newcommand{\id}{\mathds{1}}}
	{\typeout{Package dsfont.sty was not found, using alternative macros.}
         \let\mathds=\mathbb
         \newcommand{\id}{\mbox{1 \kern-.59em {\rm l}}}}

\usepackage[pdftex,bookmarksnumbered=true,breaklinks=true]{hyperref}
\usepackage[numbers,square,comma,sort&compress]{natbib}

\newcommand{\uim}{UV/IR mixing}
\newcommand{\nc}{non-com\-mu\-ta\-tive}

\newcommand{\naiv}{na\"ive}
\newcommand{\eqnref}[1]{Eqn.~(\ref{#1})}		% for equations with preceding Eqn.
\newcommand{\starco}[2]{\left[ #1\stackrel{\star}{,}#2\right] }		% star commutator
\newcommand{\staraco}[2]{\left\{ #1\stackrel{\star}{,}#2\right\} }	% star anticommutator
\newcommand{\vvar}[3]{\frac{\d^2 #1}{\d #2\d #3}}			% variational deriv. for two variables
\newcommand{\pa}{\partial}						% partial derivative sign
\newcommand{\diff}[2]{\frac{\pa #1}{\pa #2}}				% partial derivative, p1=numerator, p2=denom.
\newcommand{\ri}{{\rm i}}						% complex unit
\newcommand{\re}{{\rm e}}						% Euler e
\renewcommand{\k}{\tilde{k}}						% k tilde
\newcommand{\g}{\gamma}
\renewcommand{\d}{\delta}

\renewcommand{\th}{\theta}
\newcommand{\mth}{\theta} % abbreviation for our dimensionless theta matrix
\newcommand{\sth}{\varepsilon} % abbreviation for the dimensionfull scalar NC-parameter
\newcommand{\st}{\bar{\sigma}}

\renewcommand{\l}{\lambda}
\newcommand{\m}{\mu}
\newcommand{\n}{\nu}

\renewcommand{\r}{\rho}
\newcommand{\s}{\sigma}
\renewcommand{\t}{\tau}

\newcommand{\G}{\Gamma}

\newcommand{\inv}[1]{\frac{1}{#1}}				% inverse of something

\newcommand{\intx}{\int\! d^4x}						% 4-dim x integral
\newcommand{\nn}{\nonumber}

\newcommand{\ig}{{\rm i}g}
\title{\begin{flushright}
       \small{TUW-11-07}\\
       \small{UWThPh-2011-13}\\
       \version
       \end{flushright}
\vspace{2em}
Extended BRST formulation of a {\nc} $U_\star(1)$ gauge model}

\author{Daniel N. Blaschke\footnotemark[1]~, Thomas Garschall\footnotemark[2]~, Franz Heindl\footnotemark[2]~, 
\\Manfred Schweda\footnotemark[2]~ and Michael Wohlgenannt\footnotemark[1]}
\date{}

\begin{document}
\maketitle
\thispagestyle{empty}
\begin{center}
\renewcommand{\thefootnote}{\fnsymbol{footnote}}
\vspace{-0.3cm}\footnotemark[1]Faculty of Physics, University of Vienna\\Boltzmanngasse 5, A-1090 Vienna (Austria)\\[0.3cm]
\footnotemark[2]Institute for Theoretical Physics, Vienna University of Technology\\Wiedner Hauptstra\ss e 8-10, A-1040 Vienna (Austria)\\[0.5cm]
\ttfamily{E-mail: daniel.blaschke@univie.ac.at, garschall@hep.itp.tuwien.ac.at, franz.m.heindl@gmail.com, mschweda@tph.tuwien.ac.at, michael.wohlgenannt@univie.ac.at}
\vspace{0.5cm}
\end{center}

\vspace{1cm}
\begin{abstract}
\noindent
In this short letter, we rediscuss the model for {\nc} $U_\star(1)$ gauge theory presented in [arXiv:0912.2634] and argue that by treating the "soft-breaking terms" of that model in the realm of an extended BRST symmetry, a future renormalization proof using Algebraic Renormalization should be possible despite the fact that the {\nc} model is non-local. In fact, the non-localities could be treated in a way similar to commutative gauge theories with axial gauge fixing where certain non-local poles appear.
\end{abstract}
\vspace{1cm}
% PACS: \\
% Keywords: 

\vfill

%
%==============================================================================
\section{Introduction}
%==============================================================================
\label{sec:intro}
For a long time new types of divergences due to a mechanism referred to as {\uim}~\cite{Minwalla:1999px,Susskind:2000} have plagued quantum field theories formulated on {\nc} spaces. In fact, only some special scalar models on Euclidean $\th$-deformed spaces are known to be renormalizable to all orders of perturbation theory~\cite{Grosse:2003, Grosse:2004b,Rivasseau:2008a} --- for a review on QFTs on {\nc} $\th$-deformed spaces see~\cite{Szabo:2001,Rivasseau:2007a,Blaschke:2010kw}. However, similar successes resp. renormalization proofs for $U_\star(1)$ gauge theories\footnote{Note, that the non-commutativity of the space coordinates alters the gauge group, and hence we denote the deformed $U(1)$ group by $U_\star(1)$.} are still missing --- although several promising approaches were made in $\th$-deformed Euclidean space~\cite{Grosse:2007,Wulkenhaar:2007,Blaschke:2007b,Wallet:2008a,Blaschke:2008a,Vilar:2009,Blaschke:2009b,Blaschke:2009e}.

In this short letter, we want to rediscuss the model for {\nc} $U_\star(1)$ gauge theory presented in \cite{Blaschke:2009e}, since we believe it to be a very promising candidate for a renormalizable {\nc} gauge field theory. Although that model has also been extended to $U_\star(N)$ gauge theories in~\cite{Blaschke:2010ck}, we will for simplicity consider only the $U_\star(1)$ case here. 
The main ideas of the model \cite{Blaschke:2009e} are based on the implementation of an IR damping mechanism similar to the scalar $\inv{p^2}$-model of Gurau et al.~\cite{Rivasseau:2008a}. A series of papers~\cite{Blaschke:2008a,Blaschke:2009a,Vilar:2009,Blaschke:2009b,Blaschke:2009c,Blaschke:2009d} led to the conclusion, that the best way to make such an extension is to employ techniques known from the Gribov-Zwanziger action~\cite{Gribov:1978,Zwanziger:1989,Zwanziger:1993,Baulieu:2009} in QCD, or more precisely, to consider ``soft-breaking'' terms leading to the required damping properties of the gauge field propagator.

Here, we will treat the ``soft-breaking terms'' differently, namely in the realm of an extended BRST symmetry implying that also the new parameters characterizing this model are part of the BRST transformations. The motivation for this change is simplicity on the one hand since one no longer needs such a great number of auxiliary fields, and on the other hand an algebraic renormalization proof of the present version seems to be manageable.

Our current starting point is the action in 4-dimensional Euclidean space given in Eq.~(21) of Ref.~\cite{Blaschke:2009e}:
\begin{align}
\label{intro:action}
\nonumber
\Gamma^{(0)} & = \intx\, \bigg(
\frac 14 F_{\mu\nu} F_{\mu\nu} + \frac{\gamma^4}{4} \left( f_{\mu\nu}\frac 1{\tilde \square^2} f_{\mu\nu} 
+ 2\rho \, \tilde \partial A \frac 1{\tilde \square^2} \tilde \partial A \right)\\
&\quad + \frac{g'}{2}\Big(\! \staraco{A_\mu}{A_\nu} \frac{\tilde \partial_\mu \tilde \partial_\nu \tilde \partial_\rho}{\sth\tilde \square^2} A_\rho \Big) + s(\bar c \partial A) \bigg)\,,
\end{align}
where the auxiliary fields have been integrated out and the physical values for the additional sources 
have been inserted. 
We use the notation of \cite{Blaschke:2009e}, where $\gamma$ is a gauge parameter with mass dimension 1, denote the dimensionless coupling of the new $3A$ vertex by $g'$ and furthermore define $\rho \equiv 2\left(\sigma + \frac{\mth^2}4 \sigma^2\right)$.
In order to be self-consistent we repeat some notational facts.

We are discussing a $U_\star(1)$ gauge model on Moyal-deformed Euclidean space. Therefore, we have the usual Moyal commutator between coordinates,
\begin{align}
\starco{x_\mu}{x_\nu} = \ri \sth \mth_{\mu\nu}\,,
\end{align}
with the star product
\begin{align}
f\star g\, (x) = \re^{\ri\frac{\sth}2 \mth_{\mu\nu} \partial^x_\mu \partial^y_\nu } f(x) g(y) \Big|_{y\to x}\,.
\end{align}
The real parameter $\sth$ has mass dimension $-2$ rendering the constant antisymmetric deformation matrix $\mth_{\mu\nu}$ dimensionless,
\begin{align}
(\mth_{\mu\nu}) = \left( \begin{array}{cccc} 0 & 1 && \\ -1 & 0 && \\ && 0 & 1 \\ && -1 & 0 \end{array} \right)\,.
\end{align}
We define the following contractions with $\mth_{\mu\nu}$:
\[
\tilde v_\mu := \mth_{\mu\nu} v_\nu \,, \qquad \tilde w := \mth_{\mu\nu} w_{\mu\nu}\,.
\]
Additionally, we recall that $f_{\mu\nu}$ is just the Abelian part of the field strength for the $U_\star(1)$ gauge field $A_ \mu$:
\begin{align}
f_{\mu\nu} = \partial_\mu A_\nu - \partial_\nu A_\mu\,,
\end{align}
whereas the non-Abelian field tensor $F_{\mu\nu}$ is given as usual by 
\begin{align}
F_{\mu\nu} = \partial_\mu A_\nu - \partial_\nu A_\mu - \ig \starco{A_\mu}{A_\nu}\,.
\end{align}
The BRST transformation is denoted by $s$, and we have
\begin{align}
\nonumber
s A_\mu & = \partial_\mu c + \ig \starco{c}{A_\mu} = D_\mu c\,, \\
\nonumber
sc & = \ig c\star c\,,\\
\label{intro:BRS}
s\bar c & = B\,, \qquad s B = 0\,,\\
\nonumber
s^2 \phi & = 0\,, \quad \forall \phi\,.
\end{align}
The fermionic ghosts are denoted by $c$ and $\bar c$ and the multiplier field defining the gauge fixing by $B$.

In order to describe the non-linearities in the BRST transformations, one usually introduces unquantized external sources $\Omega^A_\mu$ and $\Omega^c$ and adds the following term to the action \eqref{intro:action}:
\begin{align}
\Gamma_{\text{ext}} = \intx \left( \Omega^A_\mu  s A_\mu + \Omega^c  sc \right)\,.
\end{align}
Neglecting the $\gamma^4$ and $g'$ dependent terms in \eqref{intro:action}, one gets the BRST-invariant action
\begin{align}
\Gamma^{(0)}_{\text{inv}} = \intx \left( \frac 14 F_{\mu\nu} F_{\mu\nu} + s(\bar c\, \partial A) + \Omega^A_\mu  s A_\mu + \Omega^c  sc \right)\,.
\end{align}
The BRST symmetry is expressed by the non-linear Slavnov-Taylor-identity:
\begin{align}
\mathcal{B}( \Gamma^{(0)}_{\text{inv}}) = \intx \left( \frac{\delta \Gamma^{(0)}_{\text{inv}}}{\delta \Omega^A_\mu} \star
\frac{\delta \Gamma^{(0)}_{\text{inv}}}{\delta A_\mu(x)} + \frac{\delta \Gamma^{(0)}_{\text{inv}}}{\delta \Omega^c} \star
\frac{\delta \Gamma^{(0)}_{\text{inv}}}{\delta c(x)} + B \star \frac{\delta \Gamma^{(0)}_{\text{inv}}}{\delta \bar c(x)}
\right) = 0\,.
\end{align}
However, the action \eqref{intro:action} violates BRST symmetry,
\begin{align}
\mathcal{B}( \Gamma^{(0)})
\ne 0\,.
\end{align}
Therefore, we want to discuss the breaking terms in the realm of an extended BRST symmetry at tree-level in order to control the perturbative results of \cite{Blaschke:2009e}.

The main feature of the action \eqref{intro:action} is the IR damping of the gauge field propagator:
\begin{align}
G^{AA}_{\mu\nu}(k) = \inv{k^2\left(1+\frac{\g^4}{(\k^2)^2}\right)}\left[\d_{\m\n}-\frac{k_\m k_\n}{k^2}-\frac{\st^4}{\st^4+k^2\left(\k^2+\frac{\g^4}{\k^2}\right)}\frac{\k_\m\k_\n}{\k^2}\right]\,,
\end{align}
with the abbreviation
\[
\st^4 \equiv  \rho\, \g^4\,.
\]

As a further comment, we should mention that the 3A vertex has the form
\begin{align}
\nonumber
\widetilde{V}^{3A}_{\r\s\t}(k_1,k_2,k_3) &= 2\ri (2\pi)^4\d^4(k_1+k_2+k_3)\Bigg(
g'\cos\left(\frac{\sth}{2} k_1 \k_2\right)\sum_{i=1}^3{\frac{\k_{i,\r}\k_{i,\s}\k_{i,\t}}{\sth (\k_i^2)^2}}\\
&\quad +g\sin\left(\frac{\sth}{2} k_1 \k_2\right)[(k_3-k_2)_\r \d_{\s\t}+(k_1-k_3)_\s \d_{\r\t}+(k_2-k_1)_\t \d_{\r\s}]
\Bigg)\,. \label{eq:3A-vertex}
\end{align}
Clearly, the action \eqref{intro:action} has the necessary structure in order to absorb the IR divergences appearing in one-loop calculations.

%
%==============================================================================
\section{Extended BRST Symmetry}
%==============================================================================
\label{sec:extended}

For the extended BRST symmetry one chooses BRST partners for the parameters $\gamma^4$ and $g'$ in the following manner:
\begin{align}
s\bar \chi = \gamma^4\,, \qquad s\bar \delta = g'\,,
\end{align}
where $\bar\chi$ and $\bar\d$ have ghost number $-1$, and mass dimensions $4$ and $1$, respectively. This leads to the following extended action:
\begin{align}
\label{extended:action}
\Gamma^{(0)}_{\textrm{inv}} & = \intx \Big( \frac 14 F_{\mu\nu} F_{\mu\nu} + s(\bar \chi \, \mathcal L^1_{\text{br}}) + s(\bar c\, \partial A) 
 + s(\bar \delta \, \mathcal L^2_{\text{br}})
+ \Omega^A_\mu  s A_\mu + \Omega^c  sc \Big)\,,
\end{align}
where
\begin{align}
\mathcal L^1_{\text{br}} & = \frac 14 \left( f_{\mu\nu}\star \frac 1{\tilde \square^2} f_{\mu\nu} 
+ 2\rho \, \tilde \partial A \star  \frac 1{\tilde \square^2} \tilde \partial A \right)
\,, \nn\\
\mathcal L^2_{\text{br}} & = \inv{2}\staraco{A_\mu}{A_\nu} \star \frac{\tilde \partial_\mu \tilde \partial_\nu \tilde \partial_\rho}{\sth\tilde \square^2} A_\rho\,.
\end{align}
The action \eqref{extended:action} is now invariant with respect to an extended BRST symmetry. In fact, the extensions have now been made BRST exact. 
In addition to the transformations~\eqref{intro:BRS}, we have
\begin{align}
\label{extended:BRS}
\nonumber
s\bar \chi & = \gamma^4\,, \qquad s \gamma^4 = 0\,,\\
s\bar \delta & = g'\,, \qquad s g' = 0\,.
\end{align} 
The symmetry of \eqref{extended:action} can now be expressed by an extended Slavnov-Taylor-identity at tree level,
\begin{align}
\label{extended:slavnov-taylor}
\mathcal{B}( \Gamma^{(0)}_{\textrm{inv}}) = &
\intx \left( \frac{\delta \Gamma^{(0)}_{\text{inv}}}{\delta \Omega^A_\mu} \star
\frac{\delta \Gamma^{(0)}_{\text{inv}}}{\delta A_\mu(x)} + \frac{\delta \Gamma^{(0)}_{\text{inv}}}{\delta \Omega^c} \star
\frac{\delta \Gamma^{(0)}_{\text{inv}}}{\delta c(x)} + B \star \frac{\delta \Gamma^{(0)}_{\text{inv}}}{\delta \bar c(x)} \right)
 + \gamma^4 \frac{\partial \Gamma^{(0)}_{\textrm{inv}}}{\partial \bar \chi} + g' \frac{\partial \Gamma^{(0)}_{\textrm{inv}}}{\partial \bar \delta}  = 0\,.
\end{align}
The physical situation is defined by $\bar \chi = 0$ and $\bar\d=0$, effectively breaking the extended BRST symmetry. 

If one uses \eqref{extended:slavnov-taylor} {\naiv}ly, one can derive the following functional equations:
\begin{align}
\label{extended:slavnov-1}
\frac{\delta^2 \mathcal{B}( \Gamma^{(0)}_{\textrm{inv}})}{\delta A_\rho(y) \delta c(z)} \Bigg|_{\phi_i=0} = - \partial_\mu^z \frac{\delta^2 \Gamma^{(0)}_{\textrm{inv}}}{\delta A_\rho(y) \delta A_\mu (z)} \Bigg|_{\phi_i=0} =0\,,
\end{align}
and
\begin{align}
\label{extended:slavnov-2} \nn
\frac{\delta^3 \mathcal{B}( \Gamma^{(0)}_{\textrm{inv}})}{\delta A_\lambda (r) \delta A_\rho(y) \delta c(z)}\Bigg|_{\phi_i=0} 
% = \frac{\delta^3 \ }{\delta A_\lambda (r) \delta A_\rho(y) \delta c(z)}\intx\!\left(\!D_\m c\!\star\!
% \frac{\delta \Gamma^{(0)}_{\text{inv}}}{\delta A_\mu}+\g^4 s\mathcal{L}_{\textrm{br}}^1+g' s\mathcal{L}_{\textrm{br}}^2\!\right)\!\!\Bigg|_{\phi_i=0} \nn\\
% &= - \partial_\mu^z \frac{\delta^3 \Gamma^{(0)}_{\textrm{inv}}}{\delta A_\lambda (r) \delta A_\rho (y) \delta A_\mu (z)}\Bigg|_{\phi_i=0} \nn\\
% & \quad + \frac{\delta^3 \ }{\delta A_\lambda (r) \delta A_\rho(y) \delta c(z)}\intx\left(\ig\starco{c}{A_\m}\star
% \frac{\delta \Gamma^{(0)}_{\text{inv}}}{\delta A_\mu}+\g^4 s\mathcal{L}_{\textrm{br}}^1+g' s\mathcal{L}_{\textrm{br}}^2\right)\Bigg|_{\phi_i=0} \nn\\
% &= - \partial_\mu^z \frac{\delta^3 \Gamma^{(0)}_{\textrm{inv}}}{\delta A_\lambda (r) \delta A_\rho (y) \delta A_\mu (z)}\Bigg|_{\phi_i=0} 
%  +\ig\left(\d_{\r\m}\starco{\d(z-y)}{\vvar{\G^{(0)}_{\textrm{inv}}}{A_\l(r)}{A_\m(z)}}+\begin{array}{c}\r\leftrightarrow\l \\ y\leftrightarrow r \end{array}\right)\Bigg|_{\phi_i=0}  \nn\\
% & \quad + \frac{\delta^3 \ }{\delta A_\lambda (r) \delta A_\rho(y) \delta c(z)}\intx\left(\g^4 s\mathcal{L}_{\textrm{br}}^1(x)+g' s\mathcal{L}_{\textrm{br}}^2(x)\right)\Bigg|_{\phi_i=0} \nn\\
 &= - \partial_\mu^z \frac{\delta^3 \Gamma^{(0)}_{\textrm{inv}}}{\delta A_\lambda (r) \delta A_\rho (y) \delta A_\mu (z)}\Bigg|_{\phi_i=0} \nn\\
&\quad +\ig\left(\d_{\r\m}\starco{\d(z-y)}{\vvar{\G^{(0)}_{\textrm{inv}}}{A_\l(r)}{A_\m(z)}}+\begin{array}{c}\r\leftrightarrow\l \\ y\leftrightarrow r \end{array}\right)\Bigg|_{\phi_i=0}  \nn\\
& \quad +\left(\g^4\diff{\ }{\bar\chi}+g'\diff{\ }{\bar\d}\right) \frac{\delta^3 \G^{(0)}_{\textrm{inv}} }{\delta A_\lambda (r) \delta A_\rho(y) \delta c(z)}\Bigg|_{\phi_i=0} 
\nn\\
&=0
\,.
\end{align}
\eqnref{extended:slavnov-1} expresses the transversality of the two-point 1PI graph at tree level, 
and \eqnref{extended:slavnov-2} yields a Ward identity for the divergence of the 3A vertex. If the model is free from anomalies, these relations hold true to all loop orders. 

At tree level, \eqnref{extended:slavnov-2} explicitly computes to
\begin{align}
&\frac{\delta^3 \mathcal{B}( \Gamma^{(0)}_{\textrm{inv}})}{\delta A_\lambda (r) \delta A_\rho(y) \delta c(z)}\Bigg|_{\phi_i=0} 
= - \partial_\mu^z \frac{\delta^3 \Gamma^{(0)}_{\textrm{inv}}}{\delta A_\lambda (r) \delta A_\rho (y) \delta A_\mu (z)}\Bigg|_{\phi_i=0} \nn\\
\nonumber
& \;
+ \ig ( \square^r \delta_{\lambda\rho} - \partial_\lambda^r \partial_\rho^r )\! \starco{\delta^{(4)}(r-z)}{\delta^{(4)}(z-y)}\! 
+ \ig ( \square^y \delta_{\lambda \rho} - \partial_\lambda^y \partial_\rho^y)\! \starco{\delta^{(4)}(y-z) }{\delta^{(4)}(z-r)}\\
\nn & \;
+ g'\staraco{\pa^z_\m\d(r-z)}{\frac{\tilde\pa^z_\m\tilde\pa^z_\l\tilde\pa^z_\r}{\sth\tilde\square_z^2}\d(y-z)} 
+ g'\staraco{\pa^z_\m\d(y-z)}{\frac{\tilde\pa^z_\m\tilde\pa^z_\l\tilde\pa^z_\r}{\sth\tilde\square_z^2}\d(r-z)} \\
& = 0 \,, \label{extended:slavnov-2b}
\end{align}
which is consistent with the explicit expression for the 3A vertex \eqnref{eq:3A-vertex}. 
The star products on the r.h.s of \eqref{extended:slavnov-2} and \eqref{extended:slavnov-2b} are with respect to $z$, and $\pa^r$ ($\square^r$ resp.) indicates that the derivative is with respect to $r$.

%
%==============================================================================
\section{Discussion of Applicability of Algebraic Renormalization}
%==============================================================================
\label{sec:discussion}

Having defined the above extended version of a {\nc} $U_\star(1)$ gauge model, one now has to study the renormalization procedure in the realm of a perturbative expansion of the model.
An according one-loop analysis has already been successfully made in~\cite{Blaschke:2009e}.

A first obstacle in performing the renormalization program to arbitrary loop order seems to be the occurrence of non-local terms at the tree-level action and in one-loop calculations. 
However, non-local terms appear also in pure commutative Yang-Mills theories quantized in non-covariant gauges, where the Leibbrandt-Mandelstam prescription \cite{Leibbrandt:1982,Mandelstam:1982,Leibbrandt:1987qv}
is used for the treatment of the unphysical poles in the gluon propagators. 
These difficulties are bypassed very elegantly in using also an extended version of the BRST symmetry, where the gauge direction $n_\mu$ (a gauge parameter) was incorporated as a BRST doublet $(n_\mu,\chi_\mu)$ \cite{Schweda-book:1998}. Having in mind the successful application of the renormalization procedure in the pure commutative Yang-Mills model, it seems possible to apply the same methods also to the {\nc} counter part presented in the sections above.

In order to carry out the algebraic renormalization one assumes that the full vertex functional $\Gamma$ can be written as a formal power series in $\hbar$ characterizing a loop expansion \cite{Schweda-book:1998,Piguet:1995}
\begin{align}
\Gamma[\phi]=\sum_{n=0}^{\infty}{\hbar^n \Gamma^{(n)}[\phi]}\,,
\end{align}
where $\phi$ stands collectively for all fields and parameters defining our model.

As explained in \cite{Piguet:1995}, one has to use the linearized BRST operator
\begin{align}
b \equiv \mathcal{B}_\Gamma = &
\intx \left( \frac{\delta \Gamma}{\delta \Omega^A_\mu}
\frac{\delta}{\delta A_\mu} + \frac{\delta \Gamma}{\delta A_\mu}
\frac{\delta}{\delta \Omega^A_\mu} + \frac{\delta \Gamma}{\delta \Omega^c}
\frac{\delta}{\delta c} + \frac{\delta \Gamma}{\delta c}
\frac{\delta}{\delta \Omega^c} + B \frac{\delta \ }{\delta \bar c} \right)
 + \gamma^4 \frac{\partial \ }{\partial \bar \chi} + g' \frac{\partial \ }{\partial \bar \delta}\,,
\end{align}
with the properties
\begin{align}
\label{linarized:properties}
\mathcal{B}_\Gamma \mathcal{B}(\Gamma) = 0 \quad \text{and} \quad \mathcal{B}_\Gamma\mathcal{B}_\Gamma = 0 \quad \text{if} \quad \mathcal{B}(\Gamma) = 0\,.
\end{align}
Under the assumption that the theory is free of anomalies, the validity of the extended BRST identity to all orders of perturbation theory is proven in a recursive way by assuming that the problem is solved to the order $(n-1)$:
\begin{align}
\nonumber
b \Gamma_{(n-1)} & = b \left( \sum_{p=0}^{n-1}{\hbar \Gamma}^{(p)} \right) = \mathcal O(\hbar^n)\\
& = \hbar^n\Delta + \mathcal O(\hbar^{n+1})\,.
\end{align}
Due to the quantum action principle, the breaking is an integrated insertion $\hbar^n \Delta \cdot \Gamma = \hbar^n\Delta + \mathcal O(\hbar^{n+1})$.

With the help of \eqref{linarized:properties} one has to solve the following consistency relation for $\Delta$ 
\begin{align}
\label{consistency_relation}
b (b \Gamma_{(n-1)}) = \hbar^n b\Delta = 0\,,
\end{align}
leading to a cohomology problem since $b$ is a nilpotent operator.

One assumes now that the general solution of \eqref{consistency_relation} is of the form
\begin{align}
\Delta = b \tilde\Delta\,.
\end{align}
If this is the case one can redefine the vertex functional in the following manner:
\begin{align}
\Gamma_{(n)} = \Gamma_{(n-1)} - \hbar^n \tilde\Delta\,,
\end{align}
so that one has
\begin{align}
b\Gamma_{(n)} = \mathcal{O}(\hbar^{n+1}) \,,
\end{align}
hence leading to the validity of the BRST symmetry at the $n$-loop level. 
If on the other hand $\Delta \neq b \tilde\Delta$, one has an anomaly discussed in \cite{Schweda-book:1998,Piguet:1995}.
This is a very brief sketch of the algebraic renormalization, but the procedure contains subtle details: Elimination of unphysical fields, parameters etc. with tricky techniques presented in great detail in the textbooks \cite{Schweda-book:1998,Piguet:1995}. 
Furthermore, this procedure allows the discussion of the physical observables of the model.

In summary, the application of the algebraic renormalization procedure could in principle be applied to our {\nc} $U_\star(1)$ model in its extended BRST version \eqref{extended:action}. 
In the UV, the physical model \eqnref{intro:action} is equivalent to the extended version \eqnref{extended:action}.
In a second step, the absence of {\uim} in higher loop corrections due to IR damping must be shown explicitly for the physical model \eqnref{intro:action}. 
In this way we hope to show renormalizability of that model.
These studies are planned to be done in a forthcoming paper.

\subsection*{Acknowledgements}
%%%%%%%%%%%%%%%%%%%%%%%%%%%%%%%

This work was supported by the Austrian Science Fund (FWF) under contracts 
P21610-N16 (D.N.B.), P20507-N16 (T.G. \& F.H.) and P20017-N16 (M.W.).

%===CONTENT ENDS HERE==========================================================

\bibliographystyle{./../../custom1}
\bibliography{./../../articles,./../../books}

\end{document}